\documentclass{article}
\usepackage{fortschritte}
\usepackage{amsmath}
\def\a{\alpha}

\def\e{\epsilon}                
\def\f{\phi}

\def\ii{{\rm i}}
\newcommand{\ex}[1]{{\rm e}^{#1}} \def\ii{{\rm i}}

\newcommand{\vet}[1]{{#1}}

\newcommand{\be}{\begin{equation}}
\newcommand{\ee}{\end{equation}}
\newcommand{\eq}[1]{(\ref{#1})}

\def\beq{\begin{equation}}                     %
\def\eeq{\end{equation}}                       %
\def\bea{\begin{eqnarray}}                     
\def\eea{\end{eqnarray}}                       
\begin {document}
\def\titleline{
\begin{flushright}
\vspace{-2cm}
{\small \rm
DFTT 22/03 \\[-.3cm]
}
\end{flushright}
\vspace{1.5cm}
%
%
%
Twisted determinants                          
and bosonic open strings
in \\ an electromagnetic field.                        
}
\def\email_speaker{
{\tt
%
%
sciuto@to.infn.it                            
}}
\def\authors{
%
%
%
%
%
Rodolfo Russo\1ad and Stefano
Sciuto\2ad\sp                                 
}
\def\addresses{
%
%
%
%
\1ad                                        %
Centre de Physique Th{\'e}orique, Ecole Polytechnique\\      %
 91128 Palaiseau Cedex, France\\            
\2ad                                        
Dipartimento di Fisica Teorica, Universit\`a di Torino\\
I.N.F.N., Sezione di Torino, Via P. Giuria 1,I-10125 Torino, Italy \\%
}
\def\abstracttext{
%
%
The bosonization equivalence between the $2$--dimensional Dirac and
Laplacian operators can be used to derive new interesting identities
involving Theta functions. We use these formulae to compute
the multiloop partition function of the bosonic open string in
presence of a constant electromagnetic field.
}
\large
\makefront
\section{Twisted determinants}

Among the basic building blocks of string amplitudes one finds the
determinants of the Laplacian and of the Dirac operator.  Since these
operators are present in the 2D action describing the free string
propagation, their determinants will appear in all perturbative
amplitudes, starting from the simplest one: the vacuum energy. A
detailed study of these determinants was performed in
\cite{Verlinde:1986kw} (and references therein).  Moreover the
equivalence between fermionic and the bosonic determinants, through
the bosonization duality, can be used as a tool for finding or proving
(in a ``physicist's way'') interesting mathematical identities among
Theta functions.

Different approaches can be used in string computations. For instance
one can use the Polyakov path integral and evaluate it directly on the
appropriate Riemann surface (see \cite{D'Hoker:1988ta} for a classical
review and~\cite{D'Hoker:2002gw} and refereces therein for the recent
applications of this technique to the two--loop case).  Another
powerful constructive way to derive the building blocks of string
amplitudes is the sewing technique. This is a very old
idea~\cite{Alessandrini:1971dd} allowing to construct higher loop
amplitudes from tree diagrams: pairs of external legs are sewn
together taking the trace over all possible states with the insertion
of a propagator that geometrically identifies the neighborhoods around
two punctures.  In this way give one gets a particular parametrization
of the $g$--loop Riemann surface known as Schottky uniformization,
where all geometrical quantities are written in terms of products over
the Schottky group (see~\cite{Russo:2003tt} and references therein).
For instance, at $1$--loop one gets the partition function of a chiral
scalar field in the form $\prod_{n=1}^\infty \left(1 -q^n \right)^3$,
where $q$ is a complex parameter representing the modulus of the
torus; it is related to period matrix entering in the corresponding
Theta--function by $q=\exp{(2\pi\ii \tau)}$, with ${\rm Im} \tau > 0
\Leftrightarrow |q| <1$.

The formulation in terms of Schottky products and the one with
Theta--functions are complementary: the geometrical expressions in
terms of Theta--functions make manifest the modular properties, while
the Schottky uniformization makes manifest the unitarity properties of
string amplitudes; in fact the Schottky multipliers $q$ are basically
the equivalent of the (exponential of the) Schwinger parameters in
field theory, but they are not suitable to deal with modular
transformations, which can be non--analytic in $q$ (for instance,
consider $\tau\to -1/\tau$).

Following the ideas of~\cite{Pezzella:1988jr}, we use the equivalence
between bosonic and fermionic systems as a device allowing to recast
Schottky products in terms of Theta--functions for any genus $g$. By
using the techniques of~\cite{DiVecchia:1988cy}, we generalize
previous analysis~\cite{Pezzella:1988jr} in two directions. On one
hand we consider fermions of spin $(\lambda, 1-\lambda)$ (the system
$(b,c)$) and the dual bosonic field $\phi$ with the background charge
$Q=1-2\lambda$, instead of focusing just on the simplest case
$\lambda=1/2$. On the other hand, for general $\lambda$ we also
consider twisted periodicity conditions along the
$b$--cycles\footnote{As usual, we call $b$--cycles the loops in the
worldsheet along the $\tau$ direction and $a$--cycles the spatial
loops along $\sigma$.}; the twisted boundary conditions along the
$b$--cycles are enforced by the insertion together with the propagator
of the operator $\ex{2\pi\ii \e j_0}$ on the fermionic side, where
$j_0$ is the fermionic number, and of the operator $\ex{2\pi\ii \e_\f
p_0}$ on the bosonic side. Notice, however, that $\e_\f = \e - 1/2$,
where the additional factor of $1/2$ is necessary in order to
reproduce in the bosonic language the usual $(-1)^F$--twist of the
fermionic traces.

This twisting can be equivalently thought as the effect of a flat
gauge connection along the $b$--cycles on a minimally coupled
fermionic system. Thus the periodicity parameters ($\e_\mu$) can be
naturally thought as non--geometrical parameters of the Riemann
surface.

The sewing technique for computing multiloop amplitudes is discussed
in detail in~\cite{DiVecchia:1988cy}. The main tool for the sewing
technique is a generalization \cite{Sciuto:1969vz}, of the usual
vertex operators, where also the emitted states are described by a
whole Hilbert space
\bea \label{V30}
\langle {\cal V}_{I}^\phi| & =& {}_{I} \langle 0, x_0=0 |
:\exp\left\{\oint_0 dz \left(-\phi^v(1-z)\partial_z
\phi_{I}(z)\right)\right\}:~,
\\ \nonumber
\langle {\cal V}_{I}^{bc} | & =& {}_{I}\langle 0; q=-Q|
: \exp\left\{\oint_0 dz \left(b^v(1-z) c_{I}(z) -
c^v(1-z) b_{I}(z) \right) \right\}:~.
\eea
Here the coordinates with the superscript $v$ describe a propagating
(virtual) string, while those with a subscript $I$ describe a generic
emitted state in the Hilbert space ${\cal H}_I$. The modes of the two
types of fields (anti)--commute among them, since they refer to
completely independent states. Let us stress that the bra--vector in
Eq.~\eq{V30} is the vacuum in the Hilbert space of the emitted string,
so that $\langle {\cal V}|$ is an operator in the $v$--Hilbert space
of the propagating string. The vertex $\langle {\cal V}|$ can be seen
as the generator of all possible three string interactions which are
obtained by saturating it with highest weight states.

Following~\cite{DiVecchia:1988cy} it is easy to construct the
generator of the $N$-point Green functions on the sphere ($g=0$),
which we write here for the $(b,c)$ system:
\beq\label{tree}
\langle V^{bc}_{N;0} | = {}_{v} \langle q=0 | \prod_{I=1}^N
\left( \gamma_I\,
\langle {\cal V}^{bc}_{I}| \gamma^{-1}_I \right) | q=0\rangle_{v}~,
\eeq
where the $N$ Hilbert spaces labeled by $I$ describe the external
fields and are all independent, while $\gamma_I$ are (arbitrary)
projective transformations mapping the interaction point from $z=1$ as
in~\eq{V30} to the arbitrary point $z_I$. The generalization $\langle
V^{bc}_{N;g} |$ of the vertex~\eq{tree} at genus $g$ is built starting
from $\langle V^{bc}_{N';0}|$ with $N'=2 g+ N$ and then identifying
$g$ pairs of Hilbert spaces (labeled by the index $\mu=1,\ldots, 2 g$)
by means of twisted propagators.  For a genus $g$ surface, the
Riemann--Roch theorem shows that the simplest non--trivial amplitude
is the correlation function with $N_b = |Q|\,(g-1)$ insertions of the
$b$ field:
\beq\label{g-l}
Z_\e^{\lambda}(z_1,\ldots,z_{N_b}) = \langle \prod_{I=1}^{N_b} b(z_I)
\rangle_{(\e,\lambda)} = \langle V^{bc}_{N_b;g} | \prod_{I=1}^{N_b}\left(
b^{(I)}_{-\lambda} | q=0\rangle_I \right) ~,
\eeq
where the subscript $\e$ reminds the non--trivial boundary conditions along
the $b$--cycles.
As it is shown in Ref.~\cite{Russo:2003tt}
one gets
\beq\label{g-l:q1}
Z_\e^{\lambda}(z_1,\ldots,z_{N_b})
\equiv {\rm det}(1-H) \;{\cal F}^{(\lambda)}
~,
\eeq
where
\begin{equation}\label{tdetH}
{\rm det}(1-H) = {\prod_{\a}}' \prod_{n=\lambda}^\infty
(1-\ex{-2\pi\ii \e \cdot N_\a} q^n_\alpha)
(1-\ex{2\pi\ii \e \cdot N_\a} q^n_\alpha)~;
\end{equation}
$N_\a$ is a vector with $g$ integer entries; the $\mu^{\rm th}$ entry
counts how many times the Schottky generator $S_\mu$ enters in the
element of the Schottky group $T_\a$, whose multiplier is $q_\a$
($S_\mu$ contributes~$1$, while $(S_\mu)^{-1}$ contributes~$-1$).  The
product ${\prod_{\a}'}$ is over the primary classes of the Schottky
group excluding the identity and counting $T_\a$ and its inverse only
once.  For the expression of ${\cal F}^{(\lambda)}$ in the general case,
see~\cite{Russo:2003tt}; in this talk we focus on the case $Q=-1$,
where the conformal weight of $b(z)$ is $\lambda=1$ and the one of $c(z)$
is zero; we consider generic values for the twist $\e_\mu$.

In the hypothesis that at least one $\e$ is non--trivial, for instance
$\e_g\not= 0$, one gets:
\beq\label{detq2}
{\cal F}^{(1)} = {\rm det}\left(
\begin{tabular}{ccc}
$\zeta_1(z_1)$ & \ldots & $\zeta_g(z_1)$ \\
\vdots &&  \vdots\\
$\zeta_1(z_{g-1})$ & \ldots & $\zeta_g(z_{g-1})$ \\
$\ex{2\pi\ii \e_1} -1 $  & \ldots & $\ex{2\pi\ii \e_g} -1 $
\end{tabular}
\right)
= \left(\ex{2\pi\ii \e_g} -1\right) {\rm det}
\left[\Omega_i(z_j)\right]~,
\eeq
where $i,j=1,\ldots,g-1$ and the expressions of the $\Omega_{\mu}(z)$
and $\zeta_{\mu}(z)$ are given in Ref.~\cite{Russo:2003tt}. From these
expressions it is possible to show that the both $\Omega_{\mu}(z)$ and
$\zeta_{\mu}(z)$ are periodic along the $a$--cycles and get a phase $
\ex{2\pi\ii \e_\nu} $ going around the cycle $b_{\nu}$. Moreover the
$\Omega$'s are completely regular on the Riemann surface, and thus
give an explicit representation for the $g-1$ twisted abelian
differentials\footnote{These $1$--forms are known as Prym
  differentials, see~\cite{Aoki:2003sy}, where the unique $\Omega$ of
  the $g=2$ case is characterized in terms of Theta functions.}. On the
other hand, for $\e_\mu\not=0$ the $\zeta_{\mu}$'s are not analytic,
but they reduce to the $g$ untwisted abelian differentials
$\omega_\mu$ when $\e \to 0$ .

In an analogous way the sewing technique allows to derive the bosonic
correlation functions corresponding to $Z_\e^{\lambda}$ of Eq.~\eq{g-l}. In
the notation of~\cite{DiVecchia:1988cy} these correlators are
\beq\label{bg-l}
Z_{\e}^{Q}(z_1,\ldots,z_{N_b}) = \langle \prod_{I=1}^{N_b}
:\ex{-\phi (z_I)}: \rangle_{(\e,Q)} = \langle V^{\phi}_{N_b;g}
| \prod_{I=1}^{N_b}\Big(| q=-1\rangle_I \Big) ~,
\eeq
where the bosonic system has background charge $Q=1-2\lambda$.

By comparing~\cite{Russo:2003tt} the bosonic and the fermionic results
one gets for $\lambda=1$ the identity
\beq\label{t2.17}
C_\e^{(1)}~ {\cal F}^{(1)} = {\cal B}_\e(z_I;\tau, \Delta,\omega)~,
\eeq
where
\bea
{\cal B}_\e(z_I;\tau, \Delta,\omega) & = &
\prod_{I=1}^{g-1} \sigma (z_I) \prod_{I<J} E(z_I,z_J)
\;\theta\left[\begin{tabular}{@{}c@{}} $0$ \\ $\e$
\end{tabular}\right]
\left(\Delta  - \sum_{I=1}^{g-1} J(z_I)
\,\Bigg|\, \tau \right)\;,
\label{t2.17b}\\
\label{Ce}
C_\e^{(1)} & = & {\prod_{\a}}' \prod_{n=1}^\infty (1- q^n_\alpha)
(1-\ex{-2\pi\ii \e \cdot N_\a} q^n_\alpha)
(1-\ex{2\pi\ii \e \cdot N_\a} q^n_\alpha)~.
\eea
The complex $g$--component vector $\vet{J}(z)$ is given by the Jacobi
map $J_\mu(z)= \frac{1}{2\pi \ii} \int_{z_0}^z \omega_\mu$ and the
explicit expressions in terms of the Schottky parametrization of
$\omega_\mu~$, $\sigma(z_I)$, of the prime form $E(z_I,z_J)$ and of
the Riemann constant $\Delta$ are given in Appendix~A
of~\cite{DiVecchia:1988cy}.

One can easily, but not trivially, check that the two sides of the
identity have the same periodicity properties; moreover, for $g=2$, we
have successfully compared the first four terms of the expansion in
powers of $q_\mu$ of the Schottky series in the two sides
of~\eq{t2.17}.

\section{Modular properties}

We now turn to the study of the modular properties of the infinite
product~\eq{Ce} and, in particular, we are interested in the modular
transformation $\tau \to (A\tau + B)(C\tau + D)^{-1}$, with $A=D=0,~
B=-C=\bf{1}$ . This map is highly non-analytic in the language of the
Schottky parametrization and can be analyzed only by using the
identity derived in the previous section. The strategy is clear: we
first rewrite the product~\eq{Ce} in terms of geometrical objects,
whose modular properties are known (see for
instance~\cite{Verlinde:1986kw},~\cite{Russo:2003tt}), then perform
explicitly the modular transformation we are interested in, and
finally use again Eq.~\eq{t2.17} to recast the result in the Schottky
language. Of course the Schottky multipliers $k_\mu$ appearing in the
final result are different from the original ones and are related to
the $q_\mu$'s in a complicated non-analytic way. So, as a first step,
we need to study the modular transformation of the ``geometrical''
expression ${\cal B}_\e$; by using the results collected in Appendix~A
of~\cite{Russo:2003tt} one gets
\begin{equation}\label{btra}
{\cal B}_\e(z_I;\tau, \Delta,\omega) =\xi\ K^{g-1}
\widetilde{\cal B}_{\tilde{\e}}(z_I;\tilde{\tau},
\tilde{\Delta},\tilde{\omega}) \;
\sqrt{\rm{det}\,\tilde\tau} ~
\ex{\ii \pi\tilde\e \tilde\tau^{-1}\tilde\e +
2\ii \pi \tilde\e \tilde\tau^{-1}\left(\tilde\Delta -
\sum_{I=1}^{g-1} \tilde{J}(z_I)\right)}~~,
\end{equation}
where $K$ is an undetermined factor independent of the $z_I$ and $\xi$
is a phase due to the quantum Weyl anomaly in two dimension. Both of
them will not be important for the final result. Finally the tilded
quantities are related to the untilded ones by

\begin{equation}\label{tut}
\tilde{\tau} = -\tau^{-1}~~,~~~
\tilde{\Delta} = \Delta\,\tilde{\tau}~~,~~~
\tilde{\omega} = \omega\,\tilde{\tau}~~,~~~
\tilde{\epsilon} = \epsilon\, \tilde{\tau}~.
\end{equation}
By using again the identity~\eq{t2.17} for $ \widetilde{\cal
B}_{\tilde{\e}}$, we can write
\begin{equation} \label{ctra}
{C}_{\e}~ {\cal F}^{(1)}_{\e} =
\xi\ K^{g-1} \,  \sqrt{\rm{det }\,\tilde\tau} ~
\ex{\ii\pi\e \tilde\tau\e + 2\ii \pi \e
\left(\tilde\Delta - \sum_{I=1}^{g-1} \tilde J(z_I)\right)}
\widetilde{C}_{\tilde{\e}}~ \widetilde{\cal F}^{(1)}_{\tilde{\e}}~,
\end{equation}
where all quantities are expressed in terms of the multipliers $q_\mu$
on the l.h.s., and in terms of the $k_\mu$'s on the r.h.s.; moreover
the r\^ole of the $a$ and the $b$--cycles is exchanged in the two
sides ($\tilde {a}_\mu=-b_\mu,\tilde {b}_\mu=a _\mu$).
By studying the $\e \to 0$ limit of this relation, one can derive the
multiloop generalization of the usual modular transformation for the
$1$--loop Dedekind $\eta$--function. In fact, at the first order in
$\epsilon$, one has $\widetilde{\cal F}^{(1)}_{\tilde{\e}} = {\cal
F}^{(1)}_{\e} \rm{det }\,\tilde\tau$, since these determinants become
linear in the $\e_\mu$'s or the $\tilde\e_\mu$'s, the $\zeta_\mu$'s
and the $\tilde\zeta_\mu$'s reduce to the usual differentials, and all
the elements are connected by Eq.~\eq{tut}.  So the determinants of
the abelian differentials on the two sides of~\eq{ctra} cancel, and
one obtains
\beq\label{geneta}
{\prod_{\a}}' \prod_{n=1}^\infty (1- q^n_\alpha)^3 = \xi\, K^{g-1}\,
({\rm{det }\,\tilde\tau})^{3/2 }
{\prod_{\a}}' \prod_{n=1}^\infty (1- k^n_\alpha)^3
\eeq
In order to avoid the appearance of undetermined factors $\xi$ and $K$
we will focus on the ratio
\begin{equation}\label{cald}
{\cal D}^{q}_\e = {\prod_{\a}}'\prod_{n=1}^\infty  \frac{
(1-\ex{-2\pi\ii \e \cdot N_\a} q^n_\alpha)
(1-\ex{2\pi\ii \e \cdot N_\a} q^n_\alpha)}
{(1- q^n_\alpha)^2}~.
\end{equation}
The modular transformation of ${\cal D}^{q}_\e$ can be again derived
from~\eq{ctra}. The additional complication brought by the presence of
a non-trivial $\e$ is that the determinants in the two sides
of~\eq{ctra} do not cancel any more. However we can simplify the
l.h.s. of the relation by integrating each $z_I$ along the $a_I$
cycle, so that the matrix defined in~\eq{detq2} becomes diagonal and
one simply gets ${\cal F}^{(1)} = \ex{2\pi\ii \e_g}-1$. On the r.h.s.,
the integration has to be taken along the $\tilde{b}_I$ cycles and is
non-trivial since the $z_I$'s appear both in the determinant
$\widetilde{\cal F}^{(1)}_{\tilde{\e}}$ and in the exponent. In this
way one obtains from Eqs.~\eq{ctra} and~\eq{geneta}
\beq\label{dqdk}
\left(\ex{2\pi\ii \e_g} -1\right) {\cal D}^{q}_\e =
{\cal D}^{k}_{\e\cdot\tilde\tau} \;
\ex{\ii \pi \e \tilde\tau \e + 2 \ii \pi \e \tilde\Delta} \;
\frac{{\rm{det}\,\rho}}{{\rm{det}\,\tilde\tau}}~,
\eeq
where $\rho_{\nu\mu}$ is a $g\times g$ matrix
\beq\label{rho}
\rho_{\nu\mu} =\frac{1}{2\pi\ii} \int\limits_{w}^{\tilde{S}_\nu(w)}
dz \left[ \tilde\zeta_\mu^{\e\cdot\tilde\tau}
(z) \ex{-\epsilon \cdot
\int\limits_{z_0}^z \tilde\omega }\right]
~~{\rm with}~\nu\not= g,~~{\rm and}~~
\rho_{g\mu} = \ex
{2\pi\ii (\epsilon\tilde\tau)_\mu} -1~;
\eeq
the tilde on $\zeta$, $S_\nu$, $\omega$ and $\tau$ indicates that
their expression is in terms of the $k_\mu$ multipliers, with twisting
parameter for the $\tilde\zeta$'s equal to
$\tilde\e=\epsilon\cdot\tilde\tau$. One can check that $\rho_{\nu
\mu}$ does not depend on the variable $w$. Analogously,
Eq.~\eq{dqdk} is independent of the base-point $z_0$ of the Jacobi
map, since the dependence on $z_0$ of $\tilde\Delta$ and of $\rho_{\nu
\mu}$ compensate each other.

This result is quite interesting for two reasons. From the
mathematical point of view, Eq.~\eq{rho} suggests how to extend
the relations presented in the previous section (in particular
Eq.~\eq{t2.17}) to the most general twists.  In fact, Eq.~\eq{t2.17}
can be generalized by replacing ${\cal B}_\e$ in Eq.~\eq{t2.17b} with
\beq\label{g1}
{\cal B}_{\e_a,\e_b} = \prod_{I=1}^{g-1} \sigma(z_I) \prod_{I<J}
E(z_I,z_J)\;\theta\left[\begin{tabular}{@{}c@{}} $-\e_a$ \\ $\e_b$
\end{tabular}\right]\left(\Delta - \sum_{I=1}^{g-1} J(z_I)\,\Bigg|\,
\tau\right)~,
\eeq
by replacing the twist $\e$ in the product~\eq{Ce} with
$\varepsilon=\e_b-\tau\cdot\e_a$ and, finally, by modifying the
determinant ${\cal F}^{(1)}$ in the following way
\beq\label{g2}
\ex{\ii\pi (\varepsilon\cdot\tau^{-1}\cdot\varepsilon
-\e_b\cdot\tau^{-1}\cdot \e_b)} ~\widehat{\cal F}^{(1)}_{\e_a,\e_b}~,
\eeq
where $\widehat{\cal F}^{(1)}_{\e_a,\e_b}$ is given by~\eq{detq2} with
the $\zeta_\mu$'s replaced by
\beq\label{g3}
\hat\zeta_\mu^{\e_a,\e_b}(z_I) = \zeta_\mu^{\varepsilon}(z_I)\exp{
\left(\frac{-2\pi\ii}{g-1} \e_a\cdot \Delta_{z_I} \right)}~.
\eeq
One can check that both ${\cal B}_{\e_a,\e_b}$~\eq{g1} and the
$\hat\zeta_\mu^{\e_a,\e_b}$'s get a phase factor of $\exp{(2\pi\ii
\e^\nu_a)}$ (or $\exp{(2\pi\ii \e^\nu_b})$) when $z_I$ goes once
around the cycle $a_\nu$ (or $b_\nu$). Therefore the
$\hat\zeta_\mu^{\e_a,\e_b}$'s in~\eq{g3} can be used as building
blocks for constructing the abelian differentials with twists both
along the $a$ and the $b$ cycles.

A physical application of the results just presented is to the
multiloop partition function of a charged open bosonic string. In
fact, by using the boundary state approach as explained
in~\cite{Frau:1997mq}, it is possible to derive the interaction
amplitude among many (bosonic) D-branes. In~\cite{Frau:1997mq} one can
find this amplitude in absence of external field (or for neutral
strings). If the ``last'' D-brane is singled out and a non trivial
gauge field strength $F_{12}=-F_{12}=f=\tan\pi\e$ is switched on, the
D-brane interaction is obviously modified. However it turns out that
this modification is quite simple: the final result for the integrand
representing the D-brane interaction is just the usual untwisted
bosonic measure (see~\cite{Frau:1997mq}) multiplied by the factor
$(\cos(\pi\e)\, {\cal D}^{q}_{\e})^{-1}$, where the twist in ${\cal
D}^{q}$ is given by the $g$-vector $\vec\e=(0,\ldots,0,\e)$. Notice
that the factor of $1/\cos(\pi\e)$ is just a rewriting of the
Born-Infeld contribution to the boundary state normalization (see for
instance Sect.~3 of~\cite{DiVecchia:1999fx}). As in~\cite{Frau:1997mq}
this result is valid for the closed string channel expression, but
thanks to the analysis of this section, it is now possible to rewrite
the same result in the open string channel. From this point of view
the interaction is just the partition function ${\cal P}_\epsilon$ for
a charged open bosonic string and becomes
\beq
{\cal P}_\epsilon = \frac{\ex{2\pi\ii \e} -1}{\cos(\pi\e)}
\left[{\cal D}^{k}_{\vec\e\cdot\tilde\tau} \right]^{-1}
\ex{-\ii \pi \vec\e \cdot\tilde\tau\cdot\vec\e - 2 \ii \pi \vec\e
 \cdot \tilde\Delta} \;
\frac{{\rm{det}\,\tilde\tau}}{{\rm{det}\,\rho}}
\; {\cal P}_0~.
\eeq
Notice that for $g=1$, the modification of the usual partition
function ({\em i.e.} the coefficient of ${\cal P}_0$) is exactly equal
to Eq.~(22) of~\cite{Bachas:bh}. In order to make the comparison
explicit one needs to change the magnetic field here introduced into
an electric one, as considered in~\cite{Bachas:bh}, ($\e\to i \e$) and
also to change the conventions on the period matrix (here purely
imaginary for open strings, $\tilde\tau \to i \tau/2$). Then it is
easy to see that the two results agree, by using $f=\tan\pi\e$ and the
explicit formula for the $g=1$ Riemann class: $\tilde\Delta= -
\tilde\tau/2 + 1/2$.
We hope to give more details on both these developments in a
subsequent publication.

\vspace{.5cm}
{\bf Acknowledgement} 
This work is supported in part by EU RTN
contract HPRN-CT-2000-00131 and by MIUR contract 2001-025249.

\end{document}